# Charge-carrier transport simulations in diamond detectors with electric-field-dependent mobility and charge-collection-distance-based trapping

Faiz Rahman Ishaqzai [a,b,*], Muhammed Deniz [b], Marta Baselga Bacardit [a], Tobias Bisanz [a], Kevin Kröninger [a], Jens Weingarten [a], Antonia Wippermann [a]

[a] AG Kröninger, Faculty of Physics, TU Dortmund University, Otto-Hahn-Str. 4a, 44227 Dortmund, Germany
[b] Department of Physics, Dokuz Eylül University, Buca, İzmir TR35160, Turkiye

## ARTICLE INFO

Dataset link: https://doi.org/10.17877/TUDODATA-2025-MIESBVSN, https://doi.org/10.17877/TUDODATA-2026-MXUECT, https://doi.org/10.17877/TUDODATA-2026-ODJ7MD, https://doi.org/10.17877/TUDODATA-2026-LS8NHQ, https://doi.org/10.17877/TUDODATA-2026-SYZQHY, https://doi.org/10.17877/TUDODATA-2026-F3XOXL

*Keywords:*
Diamond detector
Allpix squared
Charge transport
Mobility
Trapping
Charge collection distance
Transient current technique

## ABSTRACT

Diamond detectors are attractive for operation in harsh radiation environments because they combine radiation tolerance, fast signal formation, and low leakage current. Realistic detector-response simulations require an accurate description of charge-carrier mobility and trapping, which determine both signal amplitude and timing. In this work, we extend Allpix Squared, a modular end-to-end detector simulation framework, with diamond-specific transport models. The implementation includes field-dependent mobility parameterizations for electrons and holes and an effective trapping model based on the charge collection distance (CCD), providing a detector-level interface to material quality and radiation-damage measurements. The mobility description is validated in the negligible-trapping limit using single-crystalline CVD diamond by comparing simulated drift velocities and transient-current signals with published reference data. For polycrystalline CVD diamond, the CCD-based trapping model is evaluated using experimentally measured CCD values and compared with laboratory transient-current-technique waveforms. The simulations reproduce the drift-velocity behavior of the scCVD reference data and, with the measured CCD as input, the main features of the degraded transient response observed in pcCVD. The presented implementation enables detector-level studies of charge collection, pulse formation, and timing performance in diamond sensors using experimentally accessible transport and trapping parameters, and provides a practical framework for simulation-driven detector development and radiation-damage studies.

## 1. Introduction

Diamond sensors are an attractive option for particle detection in harsh operating conditions, in particular in environments combining high radiation fluence, high particle rates, and demanding timing requirements [1–3]. Their performance is governed by charge carrier mobility and charge carrier trapping, which together determine the signal size and timing response [4,5].

Accurate detector simulations are essential to translate material-level models into detector parameters or vice versa in order to support real-world detector design. Allpix Squared is a modular, end-to-end simulation framework for semiconductor tracking and vertex detectors, providing a configurable chain from energy deposition (via Geant4) to charge transport and digitized detector response [6]. It provides access to detector-level observables, e.g. the collected charge, the signal pulse shape as well as timing information, making it a well-suited tool for studying the impact of material properties on semiconductor detector operation. Its modular architecture allows the implementation of transport models and material descriptions for other semiconductors than silicon.

Diamond-detector response has previously been modeled at several levels of physical detail. Custom numerical models have described grain-dependent transport, spatially non-uniform trapping, diffusion, polarization, and induced-signal formation in pcCVD sensors [7,8]. Weightfield2 provides a dedicated two-dimensional simulation of energy deposition, carrier transport, Shockley–Ramo signal induction, and simplified electronics response for silicon and diamond sensors [9]. At the device level, TCAD-based studies have incorporated deep defect states and irradiation-induced polarization, and have coupled Geant4 energy deposition to the electrical response of specialized three-dimensional diamond sensors [10–12]. Other application-specific

* Corresponding author at: AG Kröninger, Faculty of Physics, TU Dortmund University, Otto-Hahn-Str. 4a, 44227 Dortmund, Germany.
*E-mail address:* faiz.ishaqzai@tu-dortmund.de (F.R. Ishaqzai).

Open access funding enabled and organized by Projekt DEAL.

workflows have combined Sentaurus with LTspice for intense-pulse detector–electronics simulations and Garfield++ with time-dependent weighting potentials for three-dimensional diamond sensors with resistive graphitic electrodes [13,14].

These approaches provide detailed descriptions of particular defect populations, grain structures, detector geometries, or readout systems, but many require device-specific geometries, meshes, and microscopic defect or circuit parameters. The present implementation provides a complementary detector-level approach. Literature-constrained, electric-field-dependent mobility parametrizations and a CCD-based effective trapping model are made available within the modular Allpix Squared framework. This allows measured diamond-material properties to be propagated through an event-level chain comprising energy deposition, drift, diffusion, trapping, and induced-signal formation or generic propagation. The implementation is not intended to replace microscopic TCAD or grain-resolved models, instead, it provides a reusable interface between experimentally accessible material parameters and detector observables.

In this work, we present an implementation of chemical vapor deposition (CVD) diamond as a sensor material in Allpix Squared where we place an emphasis on two ingredients required for realistic detector-response simulations. First, we introduce field-dependent mobility models for electrons and holes to describe drift velocity, impacting charge collection time, time resolution, and charge sharing. Second, we implement an effective trapping description expressed through the charge collection distance (CCD), which directly links to the collected charge and provides a convenient interface to polycrystalline structures and radiation-damage measurements.

We distinguish two practically relevant scenarios. For an unirradiated single-crystalline CVD diamond (scCVD), we validate the field-dependent mobility model in the negligible-trapping limit, where charge collection distance is consistent with the sensor thickness under typical operating voltages. For poly-crystalline CVD diamond (pcCVD), we assume the same mobility parameterization, which has been shown to be a good approximation [12] and model signal loss with an effective trapping description, quantified by supplying the charge-collection distance (CCD) as an external input.

The remainder of this paper is organized as follows. Section 2 introduces the mobility and trapping models, and describes their implementation in Allpix Squared. Section 3 summarizes the reference datasets, experimental setups and comparison procedure. Section 4 presents results for the validation of the implementation of mobility and trapping models. The study is concluded in Section 5.

## 2. Mobility and trapping models

In this section, we introduce the mobility and trapping models for diamond implemented in Allpix Squared.

Diamond is available as a material in Geant4 and can be used in Allpix Squared. The energy required to create electron–hole pairs and the Fano factor are taken from standard diamond detector literature [15]. The main missing components for a realistic detector-response simulation are diamond-specific mobility and trapping models, which we have implemented in this work.

### 2.1. Mobility models

Charge-carrier transport in Allpix Squared is simulated using a drift–diffusion model that accounts for field-dependent mobility, stochastic diffusion, and trapping. In the absence of a magnetic field, the drift velocity is given by

$$\mathbf{v}_{\mathrm{d}}(\mathbf{E}) = \mu(E)\,\mathbf{E}, \qquad E = |\mathbf{E}|, \tag{1}$$

where $\mu(E)$ is the field-dependent mobility. Diffusion is treated as a random walk, with the diffusion coefficient related to the mobility through the Einstein relation,

$$D(E) = \frac{k_{\mathrm{B}} T}{q}\,\mu(E), \tag{2}$$

where $k_{\mathrm{B}}$ is the Boltzmann constant, $T$ is the absolute temperature, and $q$ is the magnitude of the carrier charge. Although a spatially uniform electric field was used in this work, Allpix Squared provides a general propagation infrastructure that supports both uniform and non-uniform electric fields. During each integration step, the electric field is evaluated at the intermediate positions used by the numerical solver, and the corresponding mobility and drift velocity are calculated using the selected mobility parameterization. For the uniform field used here, these local evaluations return the same electric field, mobility, and deterministic drift velocity throughout the carrier trajectory. Nevertheless, propagation is performed stepwise to update the carrier position, stochastic diffusion, and trapping and, in `TransientPropagation`, to calculate the induced charge along the trajectory. For a non-uniform field description, the same procedure also accounts for the position dependence of the electric field, mobility, and drift velocity. Electrode-edge effects and other spatial field variations were not included in the present simulations.

The numerical time step depends on the propagation module. For the `GenericPropagation` simulations, the adaptive time-step scheme was used with `timestep_start`=0.01 ns, `timestep_min`=1 ps, `timestep_max`=0.5 ns, and `spatial_precision`=0.25 nm which defines the target numerical tolerance for the embedded Runge–Kutta–Fehlberg estimate of the local carrier-position integration error. For the `TransientPropagation` simulations, a fixed time step of 0.01 ns was used, which also determines the temporal granularity of the induced-charge calculation.

Based on our recent review of charge-carrier mobility in diamond, we use the Piecewise mobility parameterization (PW model) to describe the drift of electrons [4]. For holes we use the Caughey–Thomas (CT) model [16]. The mobility model can be chosen and configured via the module configuration keys. An example configuration for the diamond mobility model is provided in Appendix A.

The PW model for electrons reads

$$v_{\mathrm{d}}(E) = \begin{cases} \mu_{0,1}\, E, & E < E_{\mathrm{t}}, \\ \dfrac{v_{\mathrm{s}}\, E}{E_{\mathrm{c}}\left[1 + (E/E_{\mathrm{c}})^{\beta}\right]^{1/\beta}}, & E \geq E_{\mathrm{t}}, \end{cases} \tag{3}$$

with $E_{\mathrm{t}} = 0.08\,\mathrm{V/\mu m}$, and the CT model for holes

$$v_{\mathrm{d}}(E) = v_{\mathrm{s}} \frac{E/E_{\mathrm{c}}}{\left[1 + (E/E_{\mathrm{c}})^{\beta}\right]^{1/\beta}}. \tag{4}$$

Here $v_{\mathrm{s}}$ is the saturation velocity, $E_{\mathrm{c}}$ the characteristic field marking the onset of the transition from the linear regime to velocity saturation, and $\beta$ a shaping parameter. For $E \ll E_{\mathrm{c}}$, Eq. (4) gives $\mu_0 = v_{\mathrm{s}}/E_{\mathrm{c}} \equiv \mu_{0,2}$, whereas in Eq. (3) the low-field mobility is $\mu_{0,1}$ and $\mu_{0,2}$ is only a scale parameter of the high-field branch. The low-field mobilities are therefore $\mu_{0,\mathrm{e}} = 1880 \pm 4$ and $\mu_{0,\mathrm{h}} = 2744 \pm 9.4\,\mathrm{cm^2/Vs}$. Since $\beta$ differs for the two carriers, the saturation velocities are approached at different rates and holes drift faster than electrons throughout the detector-operation range, e.g. $v_{\mathrm{d,e}} = 64.4\,\mathrm{\mu m/ns}$ and $v_{\mathrm{d,h}} = 89.0\,\mathrm{\mu m/ns}$ at $E = 1\,\mathrm{V/\mu m}$, consistent with TCT measurements [17,18] and our review work [4].

The model parameters used in this work [4] are summarized in Table 1. The uncertainties on $E_c$ and $\beta$ are nonzero but appear as 0.00 because they were rounded to two decimal places. All mobility-model parameters were held fixed at their adopted central values in the simulations, and the quoted parameter uncertainties were not propagated to the simulated observables.

**Table 1**
Parameters of the mobility models used for charge-carrier transport in diamond in Allpix Squared adopted from [4].

| Carrier | Model | $v_s$ ($10^6$ cm/s) | $E_c$ (V/μm) | $\mu_{0,1}$ (cm²/Vs) | $\mu_{0,2} = v_s/E_c$ (cm²/Vs) | $\beta$ (–) |
|---|---|---|---|---|---|---|
| Electrons | PW | 26.59 ± 0.51 | 0.56 ± 0.00 | 1880 ± 4 | 4748 ± 92 | 0.41 ± 0.00 |
| Holes | CT | 15.09 ± 0.05 | 0.55 ± 0.00 | – | 2744 ± 9.4 | 0.88 ± 0.00 |

*Note:* Here, $v_s$ is the saturation velocity, $E_c$ the characteristic electric field, $\mu_{0,1}$ and $\mu_{0,2}$ the mobility scale parameters, and $\beta$ the parameter controlling the transition towards velocity saturation.

### 2.2. Trapping model

Charge trapping reduces the collected signal and modifies the transient current response. For diamond detectors, trapping is conveniently characterized at detector level by the charge collection distance (CCD). For a planar sensor of thickness $d$, the charge collection efficiency (CCE) is defined as

$$\mathrm{CCE} \equiv \frac{Q}{Q_0}, \qquad \mathrm{CCD} = d\,\mathrm{CCE}, \tag{5}$$

where $Q$ is the collected signal charge and $Q_0$ is the reference charge expected for complete collection of the same ionization. The second relation specifies the CCD normalization used for the traversing-particle measurements in this work. CCE therefore expresses the same charge-collection response as a dimensionless fraction.

To map CCD to an effective transport description, we use the Hecht relation [19] expressed in terms of effective carrier mean free paths (trapping lengths) $\lambda_e$ and $\lambda_h$ for electrons and holes, respectively,

$$\frac{Q}{Q_0} = \frac{\lambda_e}{d}\left[1 - \frac{\lambda_e}{d}\left(1 - e^{-d/\lambda_e}\right)\right] + \frac{\lambda_h}{d}\left[1 - \frac{\lambda_h}{d}\left(1 - e^{-d/\lambda_h}\right)\right]. \tag{6}$$

This relation preserves the experimentally relevant mapping between CCD and collected charge. Eq. (6) is the Hecht relation averaged over the generation depth, appropriate for a particle that ionizes uniformly across the thickness. For unirradiated scCVD, trapping is usually negligible, and so CCD $\approx d$. In contrast, CCD is smaller than d for pcCVD due to intrinsic defects in particular at grain boundaries and in this case Eq. (6) implies that CCD $\approx \lambda$, where $\lambda = \lambda_e + \lambda_h$. In addition, radiation damage can lead to additional sources of trapping and so the CCD can be parameterized as

$$\frac{1}{\mathrm{CCD}(\Phi)} = \frac{1}{\mathrm{CCD}_0} + k\,\Phi, \tag{7}$$

where $\mathrm{CCD}_0$ represents the unirradiated case ($\Phi = 0$) [20]. Within this effective description, the CCD of pcCVD diamond can be parameterized as a function of fluence and used as an input to simulate radiation-induced signal degradation. In the regime where charge collection is dominated by radiation damage or intrinsic defects, such as grain boundaries, the quantities CCD, effective trapping length (mean drift path), and fluence are thus related through an effective mapping,

$$\mathrm{CCD} \leftrightarrow \lambda \leftrightarrow \Phi. \tag{8}$$

In this work, signal degradation caused by any agent (defects, grain boundaries, irradiation etc.) is evaluated outside the transport code and characterized by the CCD parameter.

Allpix Squared calculates the trapping probability as an exponential decay of the simulation time-step. As our description is in terms of CCD and mean drift path ($\lambda$), we calculate the trapping probability in terms of distance-step ($\Delta$x), so that for a drift distance $\Delta$x during a step, the probability to trap within that step is

$$P_{\mathrm{trap}} = 1 - \exp\left(-\frac{\Delta x}{\lambda}\right). \tag{9}$$

Thus, trapping is integrated into stepwise propagation and affects the final charge carrier state as well as induced signals.

*Mapping CCD to $\lambda$.* The simulation accepts CCD as the primary detector-level trapping input, while Eq. (9) needs $\lambda = \lambda_e + \lambda_h$ as an input that we need to calculate from Eq. (6). Because it depends on $\lambda_e$ and $\lambda_h$ individually, we introduce a configurable ratio,

$$r \equiv \frac{\lambda_h}{\lambda_e}, \tag{10}$$

with default $r = 1.3$, adopted from [20] and with an option to override. For a given thickness $d$ and an input CCD, we solve Eq. (6) for $\lambda_e$ and then set $\lambda_h = r\lambda_e$.

The configuration of the CCD-based trapping model, including its input parameters, is illustrated in Appendix A.

## 3. Experimental setups and measurements

This section summarizes the datasets and experimental setups used to validate the charge-transport simulations.

For scCVD, we benchmark the simulated drift velocity and transient response against published reference measurements. The reference dataset comprises digitized drift-velocity data from 17 TCT experiments covering more than 45 scCVD samples over the field range 0.009–12 V/μm, obtained with both $\alpha$ and laser excitation. The numerical values were obtained by digitizing the published figures and are available in Ref. [21]. Benchmarking against this aggregated dataset rather than against a single in-house sample is deliberate. Reported scCVD transport parameters show a large sample-to-sample spread arising from crystal quality, surface termination and contacts, excitation source, and the field window probed, so the pooled data provide a more representative test of the implemented mobility models than any individual crystal would.

For pcCVD, we performed electrical (IV/CV), $\beta$-source CCD, and $^{241}$Am-TCT measurements and compared the experimentally obtained data to the pcCVD simulation results.

The pcCVD sensor used in this work is a planar diamond detector (Element Six) with sensor dimensions of $10 \times 10\,\mathrm{mm}^2$, electrode dimensions of $8 \times 8\,\mathrm{mm}^2$, and a thickness of $d = 500\,\mu\mathrm{m}$.

We distinguish between unpumped and pumped pcCVD diamond detector states. "Pumping" (also referred to as priming) denotes a pre-irradiation of the diamond with ionizing radiation prior to data taking in order to fill trapping states and stabilize the internal field and electrical response, which typically improves the charge collection efficiency. In this work, pumping was performed by exposing the detector to a $^{90}$Sr $\beta$ source for ~20 min before applying the bias voltage and the corresponding measurement series.

### 3.1. Electrical measurements

Electrical characterization (IV/CV) (supporting data [22]) is performed to define a stable bias range for subsequent CCD and TCT measurements. The current–voltage characteristics (IV curve) is measured to verify low leakage currents and the absence of early breakdown. This supports the assumption that CCD (charge integration) and TCT pulse shapes are governed by bulk transport and trapping rather than contact artefacts or baseline drift. The measurements of capacitance–voltage profiles indicate electrostatic stability (approximately bias-independent capacitance), which is consistent with the planar, near-uniform field assumption used to map $V$ to $E$ and to compare measured and simulated transients.

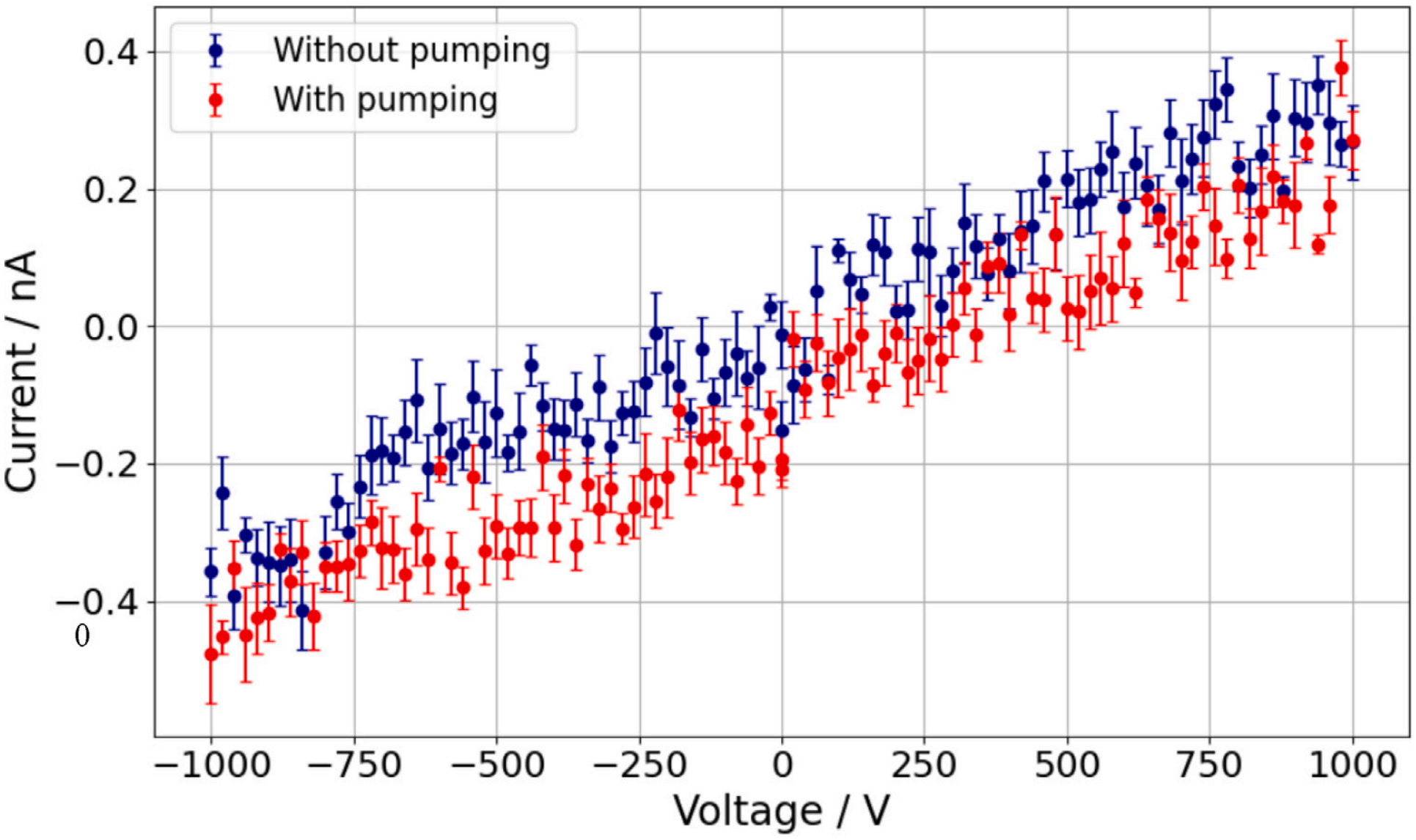

**Fig. 1.** Leakage current as a function of applied bias. Data are shown without pumping (blue) and after 20 min pumping (red).

#### 3.1.1. *Current–voltage characteristics*

Current–voltage characteristics were recorded with a Keithley 2410 sourcemeter, operated with a current compliance limit, using an MPI TS3500-SE wafer prober. The sweep was split into two unidirectional branches, from 0 V to +1000 V and from 0 V to −1000 V, in steps of 20 V, at room temperature and in the dark. At each voltage five readings were taken, the plotted value is their mean and the uncertainty their standard deviation. The same procedure was applied in the unpumped state and after 20 min of $\beta$-pumping. This protocol records a quasi-steady-state leakage current and, being unidirectional from zero bias in both branches, does not probe the hysteresis behavior reported for bidirectional sweeps of diamond sensors [23]. Fig. 1 shows the leakage current as a function of the applied bias voltage for pumped (red) and unpumped (blue) sensors. The leakage current remained below ~0.4 nA over the full bias range. It increased only slightly and approximately linearly with voltage, indicating stable, near-ohmic contact behavior. Differences between pumped and unpumped IV curves were constant and about 0.1 nA.

#### 3.1.2. *Capacitance–voltage profiles*

Capacitance–voltage profiles were recorded with an Agilent 4284A LCR meter connected through a bias box, which also provides the cable- and fixture-capacitance correction, superimposing a 20 kHz, 0.5 V AC excitation on the DC bias. The profile was measured continuously from −1000 V to +1000 V in steps of 20 V, with five readings averaged per point. As seen in Fig. 2, the capacitance is essentially voltage-independent at approximately $6.127\,\mathrm{pF}$ in the unpumped state. After pumping it is reduced by about $1.8\,\mathrm{fF}$ ($\approx 0.03\%$) and remains equally stable. The residual bias dependence of the plateau amounts to about $1\,\mathrm{fF}$ ($\approx 0.02\%$) peak-to-peak, with a shallow minimum near zero bias and a slight, approximately symmetric increase towards both polarities. This variation is of the same order as the point-to-point reproducibility of the measurement.

The first bias point deviates from the otherwise flat plateau by about $1.3\,\mathrm{fF}$, which is compatible with a settling artefact (HV/instrument stabilization and cable/fixture charging) immediately after the initial high-voltage step. A residual, order-of-magnitude smaller settling contribution persists over roughly the first ten bias points of the sweep and is visible as an increased scatter and as enlarged uncertainties in that region.

A parallel-plate estimate using $\epsilon_r \approx 5.7$, the electrode-overlap area $A = 8 \times 8\,\mathrm{mm}^2 = 0.64\,\mathrm{cm}^2$, and $d = 500\,\mu\mathrm{m}$ yields $C \approx 6.46\,\mathrm{pF}$. This is close to the measured value of $6.127\,\mathrm{pF}$, with a difference of approximately 5.2%. With the other parameters fixed at their nominal values, the measured capacitance would correspond to an effective electrode area of $60.7\,\mathrm{mm}^2$, an effective relative permittivity of 5.41, or an effective thickness of $527\,\mu\mathrm{m}$. The residual difference cannot therefore be assigned uniquely and may arise from combined uncertainties in the electrode dimensions, sensor thickness, dielectric constant, and cable/fixture correction. The approximately bias-independent capacitance demonstrates a stable terminal response over the investigated bias range, but, as a global electrical quantity, it does not by itself establish spatial uniformity of the internal electric field. Finite-electrode fringe fields may be present near the electrode perimeter, whereas the central bulk region is approximated by the planar uniform-field model. Because the non-segmented readout provides no event-by-event position information, no explicit central-region selection was applied. The comparison with simulation should therefore be interpreted as an effective bulk description.

### 3.2. *CCD measurements*

CCD measurements were performed to quantify the charge-collection performance of the pcCVD detector and to extract the detector-level input parameters required for the effective trapping model in Allpix Squared. Supporting data can be found in Ref. [24].

#### 3.2.1. *Experimental setup and readout chain*

The detector was operated in a coincidence setup with a $^{90}$Sr $\beta$ source (sealed source, activity $A$ = 37 MBq on 20.11.1995). The source is in secular equilibrium with its daughter $^{90}$Y and therefore emits a continuous $\beta^-$ spectrum with endpoint energies of $E_{\mathrm{max}}$ = 0.546 MeV ($^{90}$Sr→$^{90}$Y) and $E_{\mathrm{max}}$ = 2.28 MeV ($^{90}$Y→$^{90}$Zr). The source was placed above the diamond, while a plastic scintillator located downstream (below the diamond) served as an event trigger for traversing particles. The scintillator output was digitized as the trigger reference, and the diamond signal was processed by a spectroscopic shaping amplifier before digitization.

The diamond was read out using an ultra-low-noise, inverting CIVIDEC Cx-L Spectroscopic Amplifier [25], with an integrated shaping stage producing a Gaussian output pulse with $\sim 180\,\mathrm{ns}$ FWHM, a gain of $12.5\,\mathrm{mV/fC}$, and an integrated Bias-Tee (HV input) for detector biasing. The amplifier is specified with an equivalent noise charge of $(300 + 10/\mathrm{pF})$ electrons, and provides a linear output range of $\pm 2\,\mathrm{V}$ with a linear input range of 180 fC. An ISEG SHQ 222M high voltage power supply was used for the scintillator trigger (−1000 V), an ISEG SHR Precision DC high-voltage power supply was used for the biasing of the diamond detector, and a TENMA 72-2710 power supply was used for the amplifier.

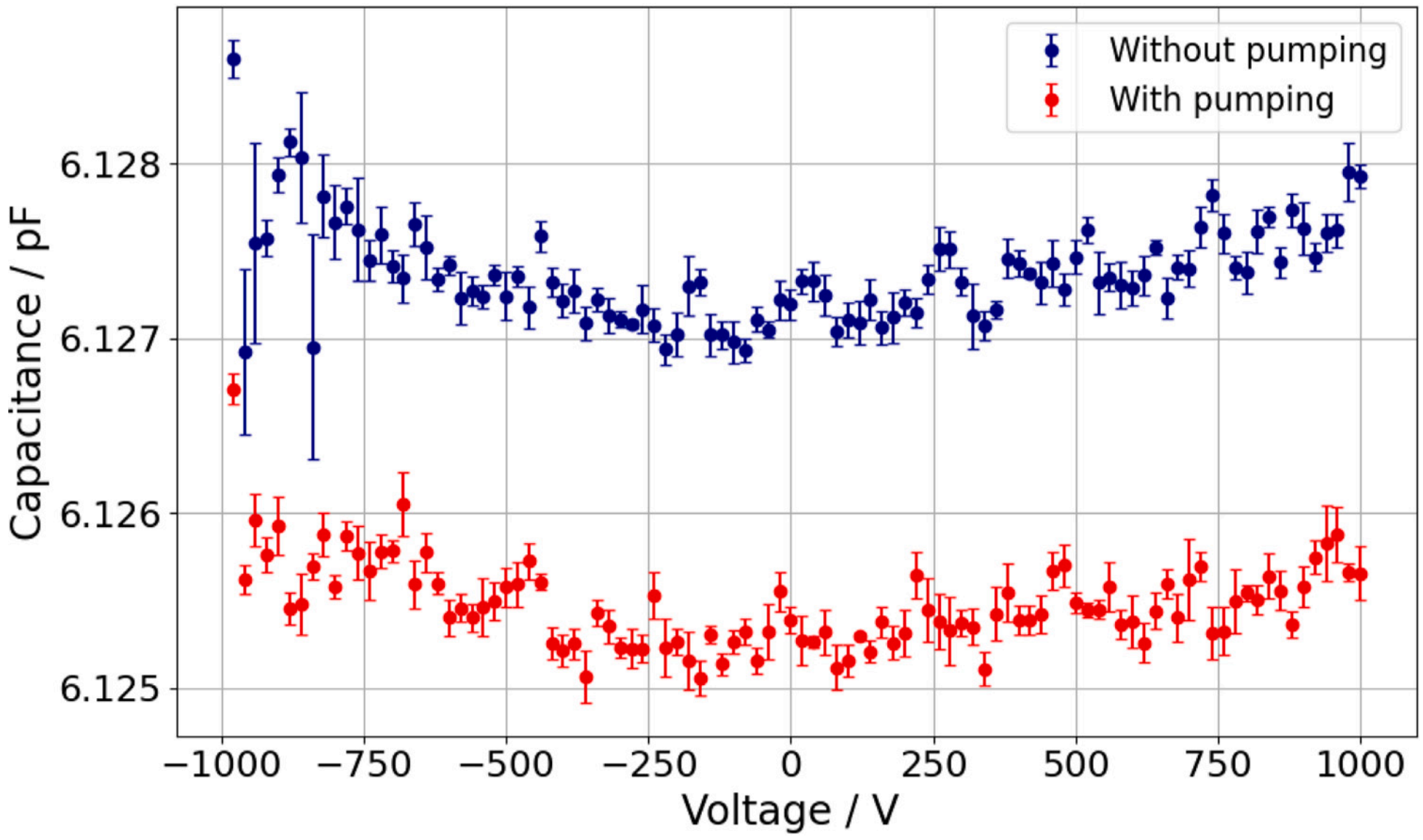

**Fig. 2.** Detector capacitance as a function of applied bias, measured from −1000 V to +1000 V. Data are shown without pumping (blue) and after 20 min pumping (red). The first bias points of the sweep are affected by instrument and high-voltage settling, which is visible as an increased scatter and enlarged uncertainties at the most negative biases.

### 3.2.2. Data acquisition

Waveforms were acquired with a Tektronix MSO5204B oscilloscope (2 GHz, 10 GS/s), controlled remotely via VISA/TCP-IP [26]. The Cx amplifier output was connected to oscilloscope channel 1 (CH1) and the scintillator trigger to channel 2 (CH2). Both channels were operated with 50 $\Omega$ input termination. The acquisition was triggered on the scintillator channel, qualified by a rectangular amplitude–time gate on the diamond channel requiring the signal to exceed 14 mV within the coincidence window.

The oscilloscope was configured for triggered waveform acquisition with a record length of $N_{\text{samples}} = 1000$ points per waveform and 2000 triggered waveforms were stored per bias setting.

### 3.2.3. Offline pulse-height (charge) reconstruction

Each stored waveform $V(t)$ is reduced to one charge estimator per event.

Baseline subtraction and noise estimation are performed in a baseline-only interval. The baseline level $b$ and the baseline noise $\sigma_{\text{base}}$ are estimated robustly from the samples in the baseline interval using the median and the median absolute deviation ($\text{MAD}_i = \text{median}\left(|V_i - b_i|\right)$),

$$b_i = \text{median}(V_i), \qquad \sigma_{i(\text{base})} = 1.4826 \times \text{MAD}_\text{i}, \tag{11}$$

evaluated over the baseline interval. The factor 1.4826 makes the MAD a consistent estimator of the standard deviation for Gaussian noise (since MAD = $0.6745\,\sigma$ for a Gaussian distribution). The median and MAD were preferred to the mean and RMS because they are less sensitive to isolated non-Gaussian excursions, such as electronic-pickup spikes or digitizer artefacts, while the scaling factor retains the conventional standard-deviation interpretation for Gaussian noise. A polarity factor $p = \pm 1$ is applied such that the expected diamond pulse polarity is mapped to a positive excursion:

$$y_{\text{eff}}(t) = p\,(V(t) - b). \tag{12}$$

The signal peak amplitude $A_{\text{sig}}$ is defined as the maximum of $y_{\text{eff}}(t)$ within a coincidence (signal) window aligned to the scintillator-trigger timing. A noise-peak estimator $A_{\text{noise}}$ is analogously defined as the maximum of $y_{\text{eff}}(t)$ in a baseline-only window centered.

### 3.2.4. Event selection (cuts)

Events are accepted if they satisfy all of the following quality criteria:

1. **SNR:** $A_{\text{sig}}/\sigma_{\text{base}} > \text{SNR}_{\text{cut}}$, with default $\text{SNR}_{\text{cut}} = 3$.
2. **Coincidence-time:** the signal peak time $t_{\text{peak}}$ must lie inside the coincidence window.
3. **Dominance:** the peak of the windowed signal must coincide with the global maximum of $y_{\text{eff}}(t)$ within a tight tolerance ($10^{-3}$ relative), rejecting events where the largest amplitude is outside the coincidence window.

For each bias setting, the number of events passing the waveform-quality criteria was recorded. Bias points with fewer than 1700 accepted events were excluded to ensure sufficient statistics for a stable pulse-height spectrum. The corresponding CCD dataset is available in the TUDOdata research data repository [24].

### 3.2.5. Charge calibration, MPV extraction, and CCD

The pulse-height is converted into an input charge using the amplifier gain $G$,

$$Q\ [\text{fC}] = \frac{A_{\text{sig[M]}}\ [\text{mV}]}{G\ [\text{mV/fC}]}. \tag{13}$$

Fig. 3 shows the charge spectrum of the accepted events, fitted with a Landau distribution convoluted with a Gaussian response function (Langau model). The most-probable value (MPV) of the fitted spectrum, $Q_{\text{MPV}}$, is used to estimate the collected charge. The separation between accepted events and baseline noise, and the effect of the SNR $> 3$ requirement, are documented in Appendix C.

The CCD is computed from $Q_{\text{MPV}}$ by converting femtocoulomb to electrons and dividing by the assumed electron–hole yield $n_{\text{eh}}$ per micrometer (default $n_{eh} = 36$ pairs/µm),

$$\text{CCD}\ [\mu\text{m}] = Q_{\text{MPV}}\ [\text{fC}] \times \frac{6241.5}{n_{\text{eh}}}. \tag{14}$$

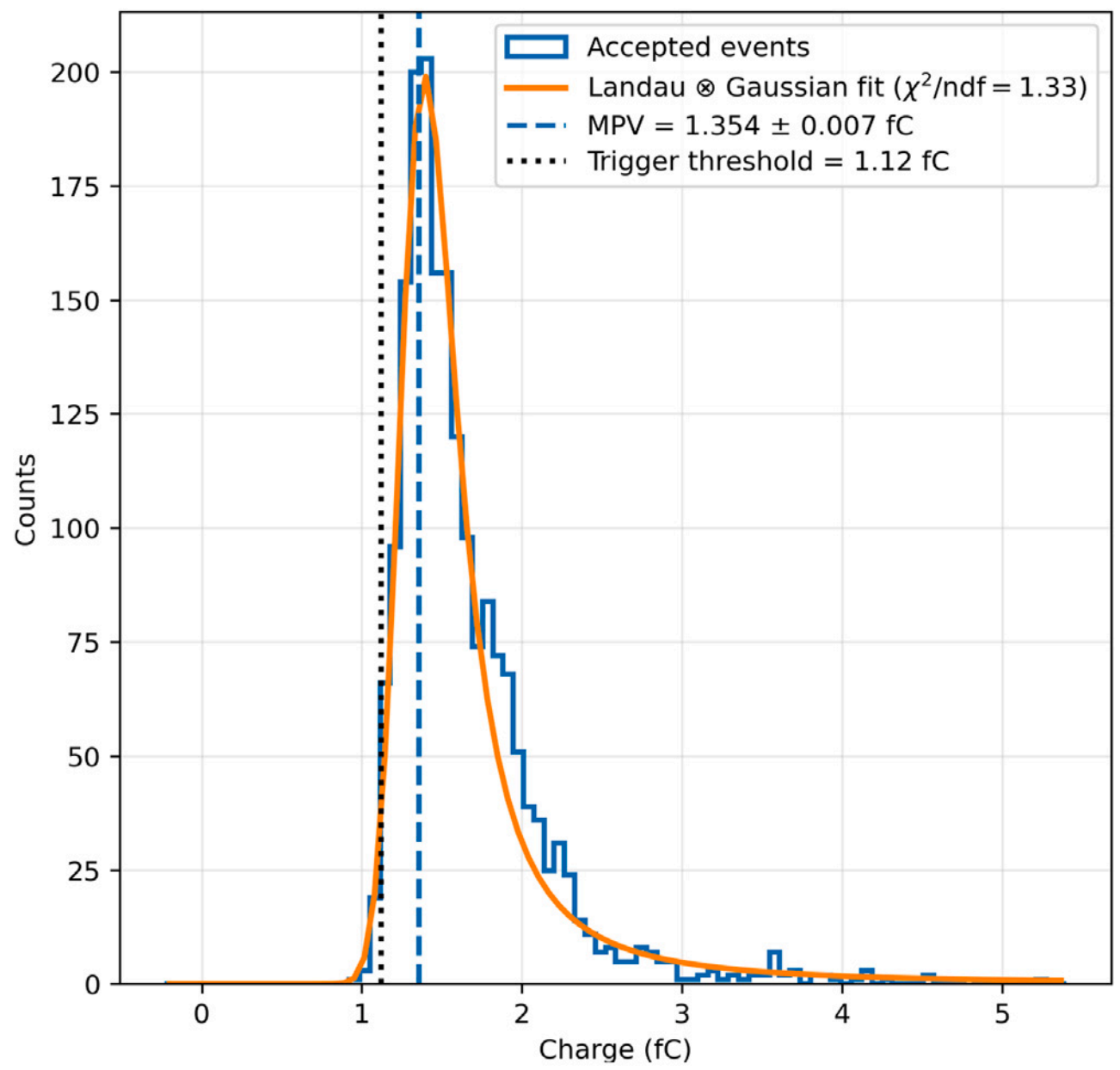

**Fig. 3.** Landau + Gaussian distribution (using Sr-90) at an electric field of 0.4 V/μm fitted with Langau model to events accepted by the three selection criteria mentioned in Section 3.2.4. The dashed vertical line (black) marks the 1.12 fC trigger threshold. The fit gives $\chi^2/\mathrm{ndf} = 1.33$.

Here, 6241.5 is the conversion factor from femtocoulomb to elementary charges i.e. 1 fC ≈ 6241.5 electrons. With the assumed electron–hole pair yield, the full-collection reference charge is $Q_0 = qn_{\mathrm{eh}}d$, where $q$ is the elementary charge. Eq. (14) therefore, corresponds to CCE = $Q_{\mathrm{MPV}}/Q_0$ and CCD = $d$ CCE.

Energy-loss fluctuations of the accepted $\beta$ particles contribute to the width and asymmetry of the measured charge spectrum. The assumed yield $n_{\mathrm{eh}} = 36$ pairs/μm is applied after extracting $Q_{\mathrm{MPV}}$ and therefore sets the absolute CCD scale without itself broadening the spectrum. The systematic uncertainty associated with this assumed yield has not been quantified and is not included in the reported statistical uncertainties.

The uncertainty on the charge collection distance ($\delta$CCD) follows from Eq. (14) by direct error propagation, since CCD is obtained from $Q_{\mathrm{MPV}}$ by multiplication with a constant factor,

$$\delta\mathrm{CCD} = \delta Q_{\mathrm{MPV}} \times \frac{6241.5}{n_{\mathrm{eh}}}, \qquad \delta\mathrm{CCE} = \frac{\delta\mathrm{CCD}}{d}. \tag{15}$$

Here $\delta Q_{\mathrm{MPV}}$ is the statistical uncertainty on the most-probable value returned by the Landau–Gaussian fit, and $\delta$CCD is the quantity shown as the error bars in Fig. 4. Systematic uncertainties (e.g. gain calibration, assumed $n_{\mathrm{eh}}$, thickness tolerance, and residual baseline systematics) are not included.

Fig. 4 shows the charge collection distance (CCD) extracted from the $^{90}$Sr spectra as a function of the absolute electric field $|E| = |V|/d$ for the two bias polarities, each measured in the unpumped state and after ~20 min of $\beta$-pumping. For all configurations, the CCD increases with $|E|$ from ~ (205–235) μm at $|E| = 0.2\,\mathrm{V}/\mu\mathrm{m}$ to a quasi-saturation regime above $|E| \gtrsim 0.6\,\mathrm{V}/\mu\mathrm{m}$, indicating reduced effective trapping at higher electric fields. Pumping improves charge collection which is consistent with trap filling and/or partial depolarization of the bulk. A systematic polarity dependence is observed over the full field range. Since the coincidence requirement selects $\beta$ particles that traverse the sensor and reach the downstream scintillator, electron–hole pairs are generated throughout the full detector thickness for both bias polarities. The observed polarity dependence therefore most likely reflects asymmetric charge transport and collection, e.g. a non-uniform distribution of trapping centers and/or carrier-dependent trapping. In our pcCVD sample, the irradiated backside corresponds to the nucleation (substrate) side, where grain boundaries and related defects are expected to be more abundant. Reversing the bias swaps which carrier species drifts towards this region. For negative bias, electrons drift towards the substrate side, whereas for positive bias, holes drift towards it. If hole trapping is stronger near the substrate side, this naturally leads to a reduced CCD for positive bias, consistent with the measured asymmetry. This interpretation is consistent with earlier work [23].

The negative-bias configuration yields a higher CCD than the positive-bias configuration by typically ~ (15–30) μm.

### 3.3. Transient Current Measurements

Transient Current Technique (TCT) measurements were performed to study the time-domain response of the diamond sensor under localized charge injection and to extract pulse-shape and timing characteristics (rise/fall times, pulse widths), which are directly dependent on the internal electric field, carrier drift, and polarization/trapping. In contrast to the $\beta$-source CCD measurements in Section 3.2, TCT uses a localized energy depositing source and records the full current transient waveform for each event and charge carrier type. The supporting data of these measurements can be found in Ref. [27].

#### 3.3.1. Experimental setup and readout chain

The same diamond detector (as in Section 3.2) was irradiated with an $^{241}$Am $\alpha$ source (activity $A$ = 5 kBq on 28.03.1996). The detector was illuminated from the back side (grounded electrode), while the front side was connected to the HV supply and read out via a broadband TCT amplifier. This geometry ensures that the charge deposition occurs close to the grounded electrode. Depending on the chosen field direction, one carrier species traverses most of the sensor thickness while the other is collected close to the generation side, enabling carrier-sensitive pulse-shape studies.

The readout used a CIVIDEC C2-TCT broadband current amplifier [28] with integrated Bias-Tee for simultaneous biasing and signal extraction. The amplifier provides 40 dB gain, a bandwidth of 0.000 01 GHz to 2 GHz, and is optimized for TCT transients from the ns range up to 50 μs. The amplifier output was digitized by a Tektronix MSO5204B oscilloscope.

#### 3.3.2. Data acquisition

Waveforms were acquired with a Tektronix MSO5204B oscilloscope controlled remotely via VISA/TCP-IP using a dedicated Python acquisition script, with 50 $\Omega$ termination. The trigger levels (±14 mV (unpumped), ±18 mV (pumped)), record length (1000), sampling rate (10 GS/s), and number of repetitions (2000) were set via an external configuration file read by the acquisition script. For each triggered event, the full raw waveform $V(t)$ was stored.

#### 3.3.3. Event selection

Offline processing and selection were performed by estimating a baseline level $b_i$ and baseline noise $\sigma_i$, for each waveform, in a baseline-only interval using robust estimators (Eq. (11)). The waveform is baseline-subtracted and automatically sign-flipped such that the transient pulse is positive,

$$V_i^{\mathrm{corr}}(t) = p\,(V_i(t) - b_i), \qquad p = \pm 1, \tag{16}$$

where $p$ is chosen from the population statistics of the individual waveform polarities.

A signal window $t \in [9\,\mathrm{ns}, 17\,\mathrm{ns}]$, a baseline window ending at 4 ns, and a noise-charge window ($[0\,\mathrm{ns}, 8\,\mathrm{ns}]$) were defined with automatic relocation to avoid overlap with the signal window. An optional late-time "ringing" window (default $[20\,\mathrm{ns}, 25\,\mathrm{ns}]$) can be used to quantify pickup/ringing.

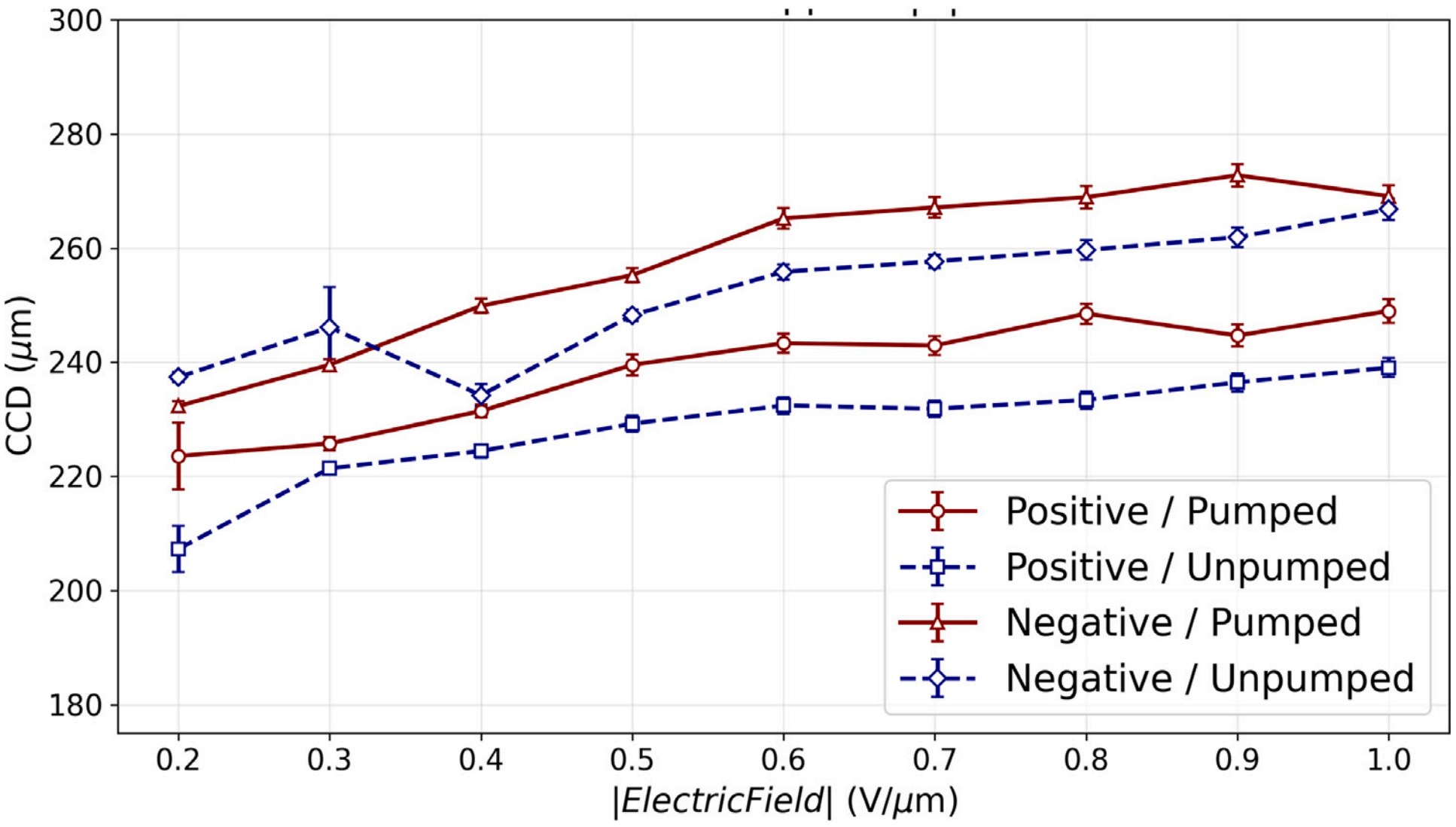

**Fig. 4.** CCD of pcCVD diamond, extracted from $^{90}$Sr spectra as a function of $|E| = |V|/d$. Results are shown for both bias polarities in the unpumped state (blue) and after ∼20 min $\beta$-pumping (red). Error bars denote the statistical uncertainty of the MPV from the Landau–Gaussian (Langau) fit.

For each event, the script defines a positive-only integrated charge in the signal window,

$$Q_i^{\mathrm{sig}} = \int_{\mathrm{sig}} \max\left(V_i^{\mathrm{corr}}(t), 0\right)\, \mathrm{d}t, \tag{17}$$

and an analogous quantity $Q_i^{\mathrm{noise}}$ in the noise window. The noise scale $\sigma_Q$ is estimated robustly from the $Q^{\mathrm{noise}}$ distribution (MAD), and the primary selection metric is

$$\mathrm{SNR}_Q = \frac{Q^{\mathrm{sig}} - \mathrm{median}(Q^{\mathrm{noise}})}{\sigma_Q}. \tag{18}$$

The default requirement is $\mathrm{SNR}_Q > 3$. Events are accepted based on a combination of (i) the $\mathrm{SNR}_Q$ requirement, (ii) a dominance cut requiring that the maximum excursion in an early-time search region coincides with the signal-window peak (suppressing late ringing), and (iii) basic finiteness/positivity checks so that waveforms are rejected if the extracted parameters are numerically broken or have impossible sign/value.

Additional cuts include an amplitude-based SNR ($A_{\mathrm{sig}}/\sigma_i$, where $A_{sig} = max V_i^{corr(t)}$), a template correlation coefficient $\rho$, an effective-width outlier rejection using $R_{\mathrm{eff}} = Q^{\mathrm{sig}}/A_{\mathrm{sig}}$, and an optional ringing metric based on the late-time RMS relative to baseline noise.

For visualization and mean-waveform stability, narrowband pickup sinusoids were subtracted using only pulse-free regions (baseline + late tail). This is performed after the event selection and therefore does not change the acceptance.

### 3.3.4. *Mean waveform and transient times*

Timing observables are extracted from the mean waveform of the accepted event set,

$$\overline{V}(t) = \frac{1}{N} \sum_{i \in \mathrm{good}} V_i^{\mathrm{corr}}(t), \tag{19}$$

and both the event-to-event standard deviation band and the standard error of the mean (SEM) are produced for diagnostics. Several parameters are extracted from $\overline{V}$: (i) peak time $t_{\mathrm{peak}}$ and amplitude $A_{\mathrm{peak}}$ (peak constrained to the signal window), (ii) FWHM from the 50% level crossings, (iii) rise time $t_{90}-t_{10}$ and fall time $t_{10}-t_{90}$ from 10% and 90% crossings, and (iv) an edge-defined pulse width using a configurable edge threshold

$$V_{\mathrm{edge}} = \max\left(f_{\mathrm{edge}}\, A_{\mathrm{peak}},\; k_\sigma\, \sigma_{\mathrm{base}}\right), \tag{20}$$

where $f_{\mathrm{edge}}$ and $k_\sigma$ are user parameters (defaults: $f_{\mathrm{edge}} = 0.10$, $k_\sigma = 0$).

For the accepted digitized waveforms, an additional pulse-width observable is derived from the analytic-signal envelope. For the mean waveform $x[n]$, the discrete analytic signal is constructed using an FFT-based Hilbert transform, and the envelope is defined as $a[n] = |z[n]|$. The corresponding envelope FWHM is then obtained as the time interval between the rising and falling half-maximum crossings of $a[n]$, determined by linear interpolation between adjacent samples. Compared with the direct waveform FWHM, this observable is less sensitive to small oscillatory structures in the transient. The extracted transient signals and the corresponding timing observables are shown in Fig. 5. The same envelope-FWHM procedure is also applied on an event-by-event basis for accepted events only and retained as a quality-control observable. For each bias setting, the analysis produces a compact set of summary and diagnostic output files.

*Relation to Shockley–Ramo theorem and weighting field.* The measured TCT transient is interpreted in terms of the induced current on the readout electrode via the Ramo–Shockley theorem. For a moving charge $q$ with drift velocity $\mathbf{v}(t)$, the instantaneous induced current is

$$i(t) = q\, \mathbf{v}(t) \cdot \mathbf{E}_w(\mathbf{x}(t)), \tag{21}$$

where $\mathbf{E}_w$ is the *weighting field* obtained by setting the readout electrode to unit potential, all other electrodes to zero, and neglecting space charge (pure electrostatics). For an ideal planar pad geometry, $\mathbf{E}_w$ is approximately uniform with magnitude $|\mathbf{E}_w| \simeq 1/d$, such that the current waveform is directly proportional to the carrier drift velocity (and thus to the local electric field through $\mathbf{v} = \mu\mathbf{E}$ in the non-saturated regime). The time integral of the transient yields the induced charge,

$$Q_{\mathrm{ind}} = \int i(t)\, \mathrm{d}t = q\, \Delta\phi_w, \tag{22}$$

where $\Delta\phi_w$ is the change in weighting potential along the carrier trajectory; in the ideal planar case, carriers traversing the full thickness contribute approximately $q$ to the integral. Deviations of the pulse shape from the ideal rectangular/triangular forms—as well as polarity/illumination-side dependencies—therefore reflect non-uniform internal fields (e.g. polarization), carrier trapping/detrapping, and velocity saturation, and provide a basis for comparison with simulated transients computed for a given electric-field and weighting-field model.

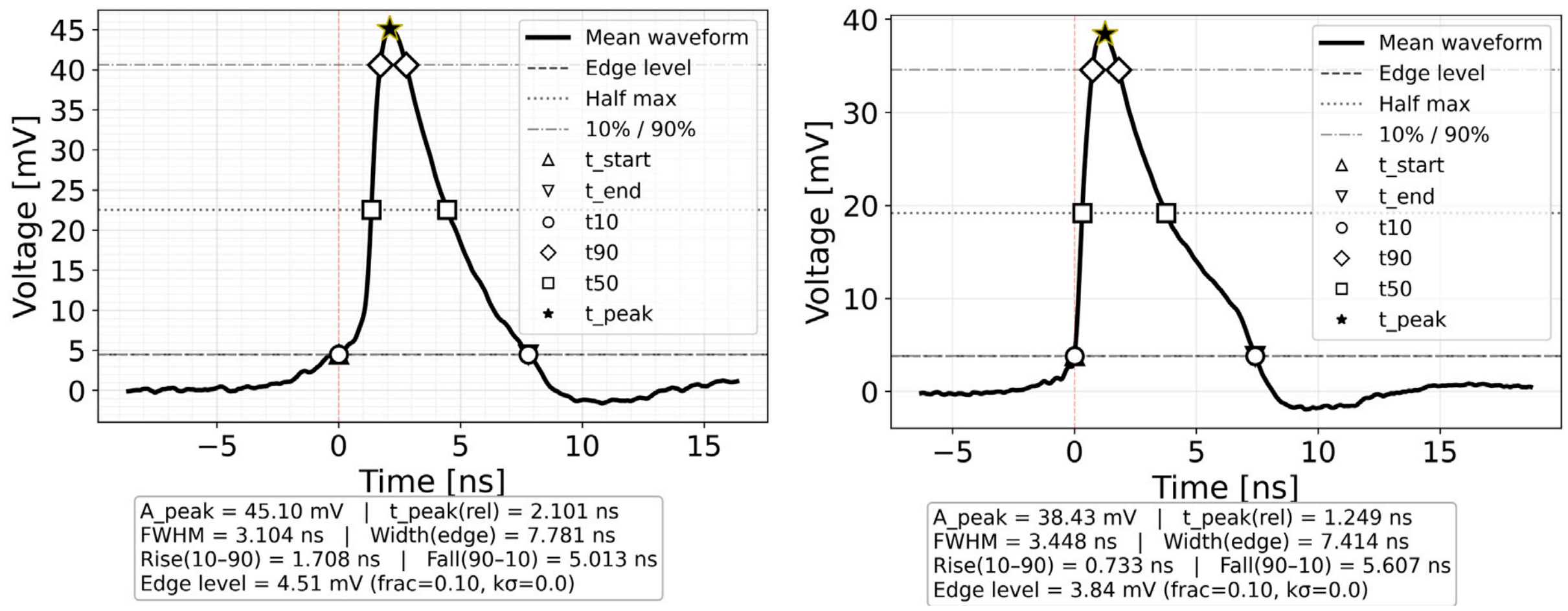

(a) (b)

**Fig. 5.** Mean measured transient current waveforms at $E = 1\,\mathrm{V}/\mu\mathrm{m}$ for (a) electrons and (b) holes. The markers indicate the characteristic timing points used in the waveform analysis, namely the 10%, 50%, and 90% level crossings on the leading and trailing edges, together with the peak position. The corresponding values of the peak amplitude, rise time, fall time, full width at half maximum (FWHM), and edge-defined pulse width are reported in each panel.

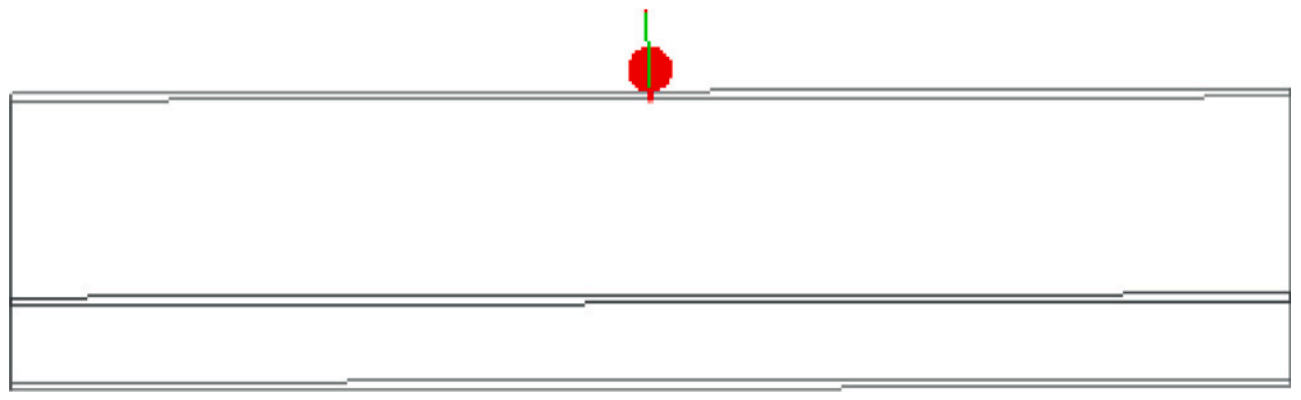

**Fig. 6.** Schematic of the simulated diamond sensor geometry, indicating the sensor thickness and the position of the Am-241 source. The red marker denotes the source location.

## 4. Simulations and results

We have simulated charge-carrier drift in diamond using a single-pixel geometry in Allpix Squared by defining a diamond sensor with variable dimensions (default: $4 \times 4 \times 0.5$ mm$^3$), as shown in Fig. 6. No explicit contacts were defined. An $^{241}$Am $\alpha$ point source was placed 1 μm above the top surface of the diamond. The supporting data for simulations is available in Ref. [29,30].

To verify that the simulated charge-carrier polarity assignment (electrons or holes) is consistent with the chosen simulation setup, we examined the transport profiles at E = 0.18 V/μm, shown in Fig. 7. As expected for scCVD with $\lambda \geq$ CCD, the drift (forward spread) dominates the diffusion (lateral spread) and full charge is collected. The Alpha particle emitted by Am-241, deposits all its energy in the first 10 μm of the sensor thickness, hence only one kind of charge carrier traverse the full thickness of the sensor.

### 4.1. scCVD mobility validation

We validated the scCVD drift velocity and transient response against published reference measurements [4,12,21,31–42].

#### 4.1.1. Electrons (PW model)

The PW model was validated in two different ways, namely the time of arrival and the transient pulses. The two approaches complement each other and both support the correct implementation of the model.

*Time of arrival.* Using the `GenericPropagation` module, charge-carrier drift was simulated for 60 bias-voltage values ranging from −1 V to −40 000 V. For this polarity of the bias voltage, only electrons traversed the full thickness of the diamond. For a sensor thickness $d$ and an extracted drift time (time of arrival) $t(E)$, the drift velocity was calculated as

$$v_{\mathrm{d}}(E) = \frac{d}{t(E)}. \tag{23}$$

Fig. 8(a) shows the extracted $v_{\mathrm{d}}(E)$ from the simulation together with digitized literature data. Relative residuals are also shown to quantify the agreement. The simulated results agree well with the literature data.

*Transient pulses and transient time.* Using the `TransientPropagation` module, charge-carrier drift was simulated and transient pulses were extracted from the simulation output for different electric fields. The transient time was estimated from the full width at half maximum (FWHM) of the corresponding pulses, as shown in Fig. 11(a) in Appendix B. The drift velocity was then calculated for the respective electric fields using Eq. (23). Fig. 8(a) shows the resulting $v_{\mathrm{d}}(E)$ dependence obtained from the transient-pulse simulations together with digitized literature data. Relative residuals are included to quantify the level of agreement. Overall, the simulated drift velocities are consistent with the literature data within the investigated field range.

The simulated transient pulse was subsequently processed with an amplifier model using a gain of 53 dB and a bandwidth of 0.01–2 GHz, consistent with the characteristics reported in Ref. [43], and employed in the measurements of Ref. [12]. The amplified simulated transients were then compared with experimentally measured transient pulses from a 526 μm-thick scCVD diamond sample, as shown in Fig. 9(a). The simulated transients exhibit a steep rise within the first nanosecond, followed by an approximately constant plateau and a rapid fall-off at the end of the signal. The experimentally reported transient pulses show broader rising and falling edges, reflecting the finite temporal response of the experimental setup, which is not fully included in the simulation. Overall, the simulations reproduce the main features of the measured waveforms and capture the observed pulse evolution. The signals shown correspond to an electric field of 0.18 V/μm.

The negative excursions following the main pulse in the measured reference waveforms are consistent with the finite response of the measurement chain, including possible baseline recovery, ringing, or cable and impedance effects. Because the complete detector–amplifier–cable–oscilloscope transfer function for the published reference data is not available, the contribution of an individual component cannot be identified conclusively. These late-time electronics

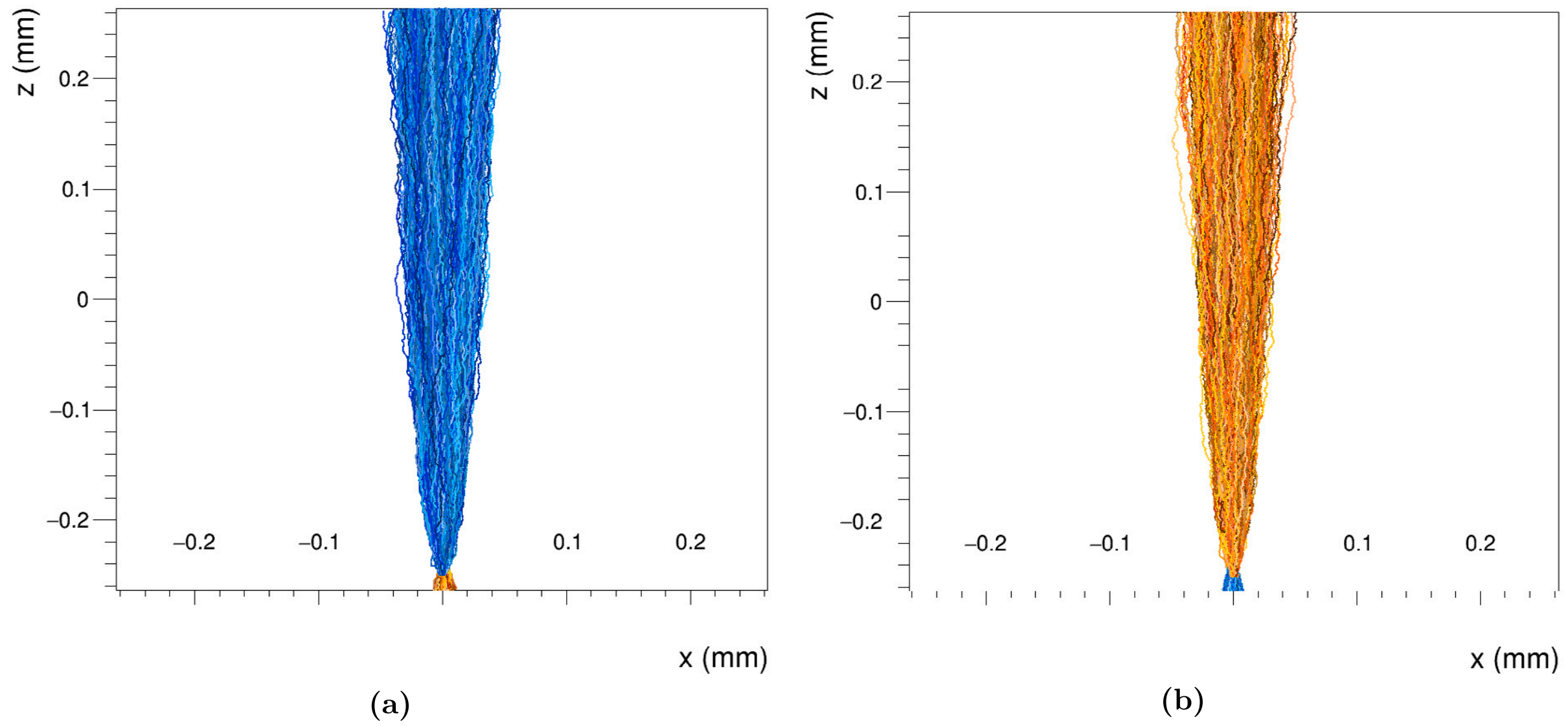

**Fig. 7.** Representative simulated charge-carrier transport trajectories of (a) electrons and (b) holes at $E = 0.18\,\mathrm{V/\mu m}$. The plots show the projected transport paths and their lateral spread during drift through the sensor bulk. The incident Am-241 (bottom) deposits all its energy within few micrometers of the sensor. The horizontal axis is not to scale and does not correspond to the physical lateral dimension of the sensor.

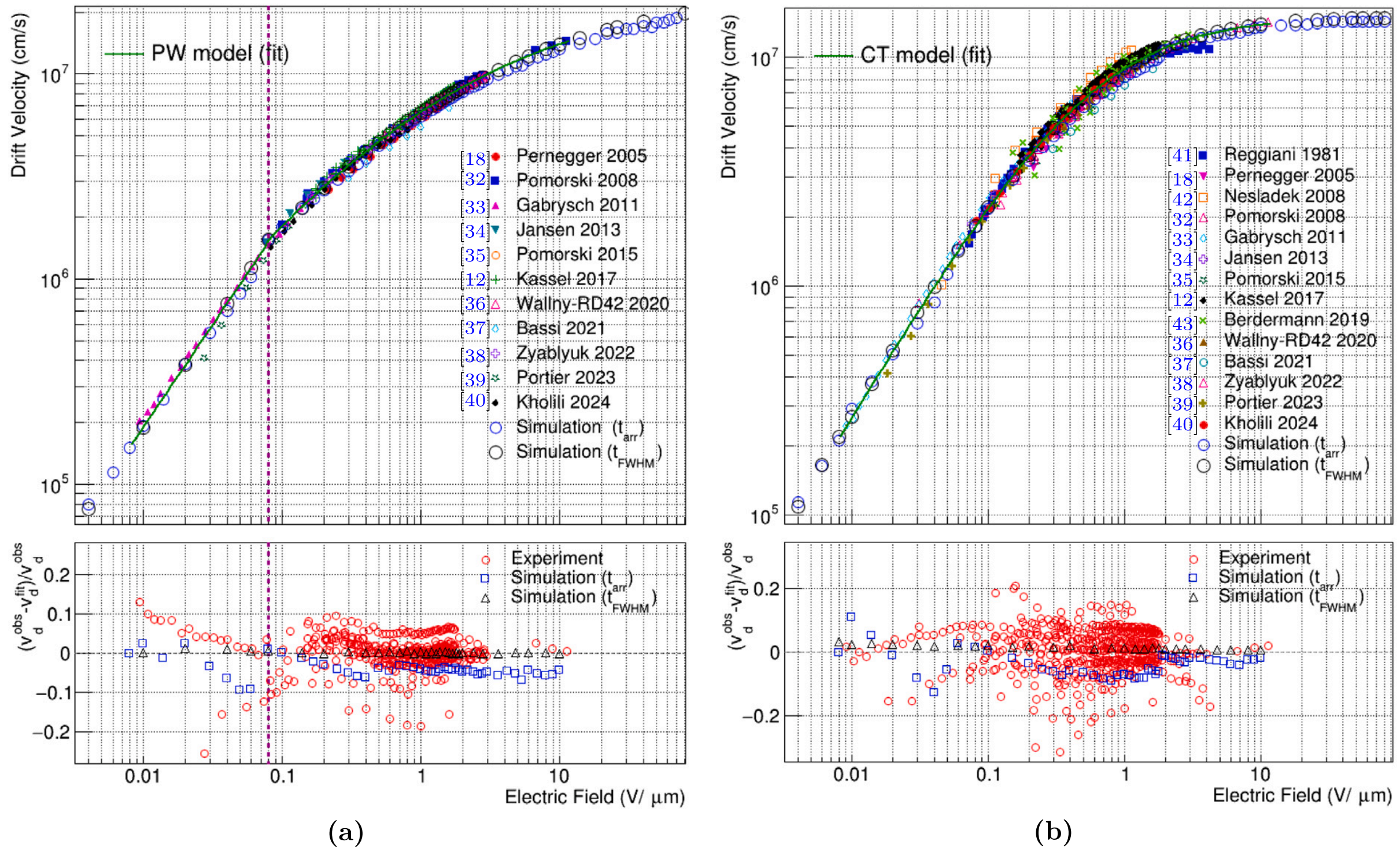

**Fig. 8.** Drift velocity as a function of electric field for (a) electrons and (b) holes, comparing simulation results with experimental data. The green line shows the fitted model obtained from the experimental data and corresponding to the parameterization used in the simulation. It is included to facilitate comparison of the simulation results with the model prediction. The lower panels show the relative residuals, $(v_d^{\mathrm{obs}} - v_d^{\mathrm{fit}})/v_d^{\mathrm{obs}}$, as a function of $E$, highlighting the agreement between the simulation and the data used to parameterize the models implemented in the simulation framework. The scatter in the "Simulation ($t_{\mathrm{arr}}$)" points reflects the uncertainty in the determination of the mean time of arrival.

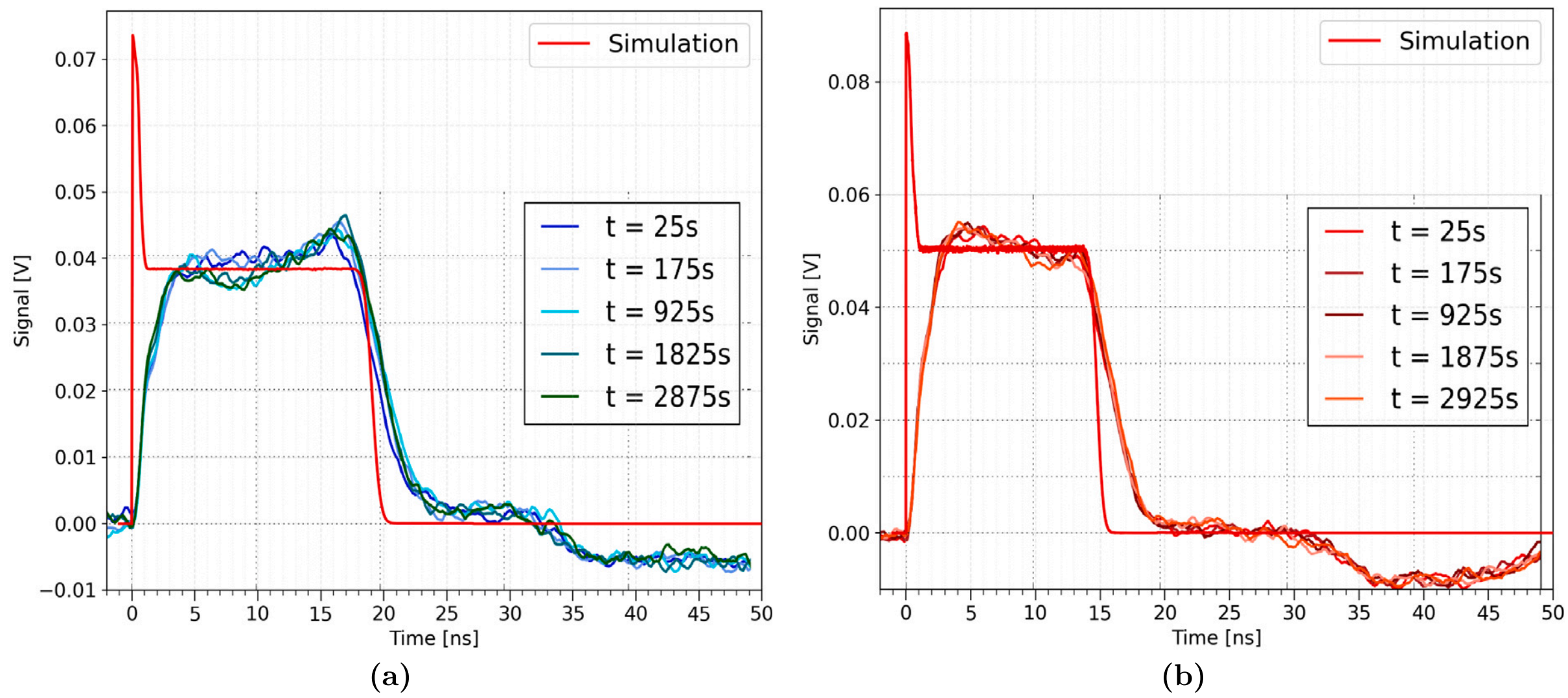

**Fig. 9.** Comparison of simulated (red) and experimentally measured TCT signals in scCVD diamond at $E = 0.18\,\mathrm{V/\mu m}$ for (a) electrons and (b) holes, using the experimental data from Ref. [12]. The color scale represents the evolution of the measured signal shape with polarization time. Polarization effects are not implemented in the present work.

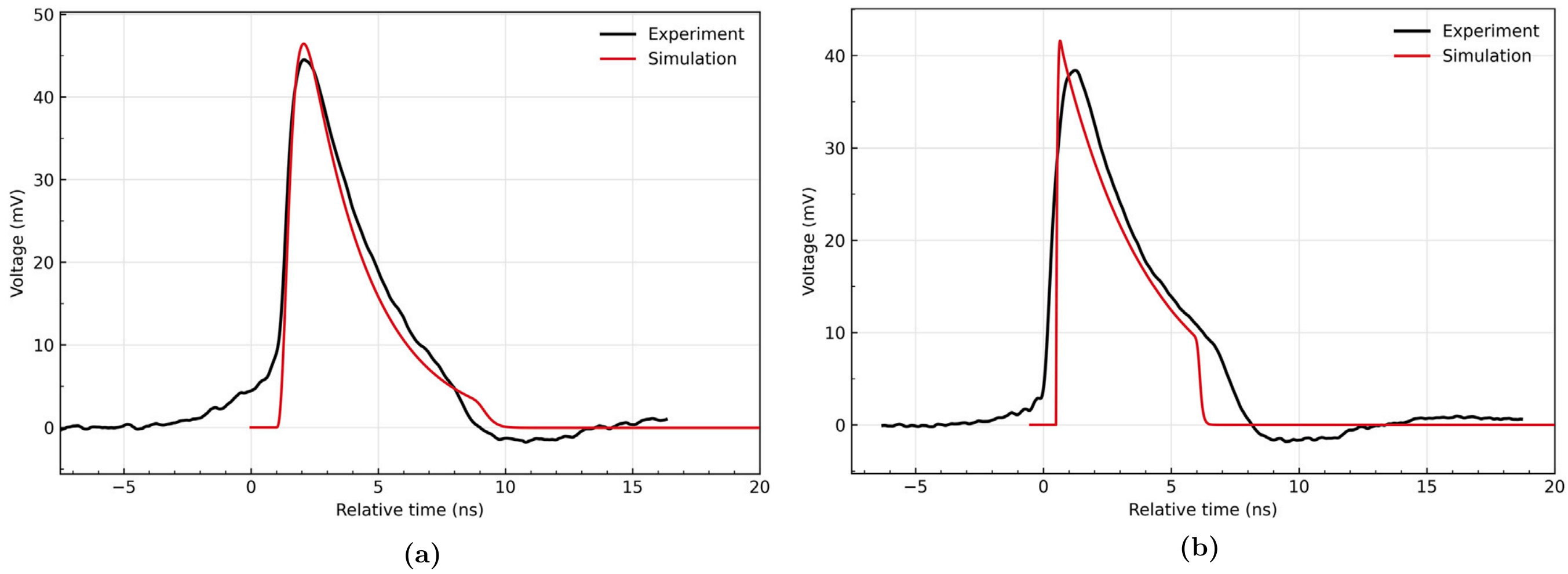

**Fig. 10.** Comparison of simulated (red) and experimentally measured (colored) TCT waveforms in pcCVD diamond at $E = 1\,\mathrm{V/\mu m}$ for (a) electrons and (b) holes. The red curves represent the post-processed simulation output and the black curves the mean measured waveforms. The indicated timing markers are derived from the measured signal and illustrate the leading-edge development, peak position, and trailing-edge evolution used for the waveform comparison.

features are therefore not included in the transport interpretation of the simulated signals.

#### 4.1.2. Holes (CT model)

The implementation of the CT model for holes was validated by following the same approaches described in Section 4.1.1. The corresponding simulation results are shown in Fig. 8(b), Fig. 11(b) in Appendix B, and Fig. 9(b), and they are in good agreement with the experimentally measured data.

### 4.2. pcCVD CCD/trapping validation

We have validated the implementation of the trapping model to reproduce measured TCT pulse shapes by comparing simulated transients with laboratory waveforms obtained using an Am-241 source (see Section 3.3). In this validation, the experimentally determined CCD values (at E = 1 V/μm) from Fig. 4 were used as inputs to the trapping model. The simulated transients were subsequently processed with the same signal-treatment procedure adopted for the experimental analysis, including the effective gain corresponding to the CIVIDEC TCT readout chain [28], such that the comparison is performed at the level of the reconstructed waveform observables.

The gain values used in the scCVD and pcCVD waveform comparisons refer to different datasets and readout chains. The 53 dB gain applies to the Particulars amplifier used for the published scCVD measurements of Kassel et al. [12]. The pcCVD measurements reported here were performed using the CIVIDEC C2-TCT amplifier, whose nominal gain is 40 dB. In the waveform analysis, an effective gain of 43 dB was used.

Fig. 10 shows the resulting comparison at $E = 1\,\mathrm{V/\mu m}$ for electron and hole drift. In both cases, the simulation reproduces the principal pulse amplitude scale, the time of the maximum, and the overall signal duration.

The two carrier waveforms are physically distinct despite their similar overall appearance. The hole waveform shows a sharp falling-edge cusp, consistent with a population of holes reaching the collecting electrode. The electron waveform has lower amplitude and a broader decay. In the present distance-based trapping model, this difference is governed by the effective trapping lengths and carrier paths rather than by drift velocity alone. In particular, the adopted ratio $\lambda_h/\lambda_e = 1.3$ gives electrons a shorter effective trapping length and therefore stronger charge loss for comparable drift paths. The lower electron velocity changes the temporal scale and can make the loss appear more extended in time but it also increases the probability of trapping. In the measured pcCVD waveform, possible trapping/detrapping dynamics, the frequency-dependent readout response, and residual pickup or noise can further broaden the signal or obscure a cusp. The present implementation models effective trapping and does not separately resolve detrapping.

For the electron signal, the simulated waveform shown in Fig. 10(a) is post-processed prior to comparison with the measured signals. In particular, the simulated transients are filtered to account for the finite bandwidth of the readout chain, and an additional smoothing/superposition procedure is applied in order to approximate the experimentally observed signal shape. The post-processed simulation remains characterized by a steeper leading edge than observed in the measured mean waveform, indicating that the initial rise of the induced signal is still not fully reproduced. In contrast, the agreement on the trailing edge is noticeably better. The simulated decay follows the measured waveform over most of the falling part of the pulse, including the slower decrease at low amplitudes. A residual difference is visible in the late-time structure, where the simulated waveform exhibits a cusp-like feature below the 10% level that is not clearly resolved in the experimental waveform.

The localized energy deposition of the $^{241}$Am $\alpha$ particles produces a dense electron–hole cloud near the sensor surface. During the initial separation and expansion of this cloud, its space charge may transiently screen and redistribute the applied electric field, potentially delaying and broadening the leading edge of the measured transient [4,44,45]. This collective evolution is expected to occur mainly on a sub-nanosecond timescale and to become less visible at higher fields because the carriers separate more rapidly. It may therefore contribute more strongly to the leading-edge broadening at lower fields, particularly for electrons. However, this contribution cannot be separated conclusively from the effects of finite readout bandwidth, diffusion, trapping, and field inhomogeneities in pcCVD material, and it is not explicitly included in the present simulation. The CCD values used as trapping inputs were obtained independently from the $\beta$-source measurements, for which the ionization is distributed along the particle track rather than concentrated within a dense surface-localized cloud.

For the hole signal in Fig. 10(b), the agreement between simulation and measurement is generally satisfactory, with the pulse duration reproduced to reasonable accuracy. In this case, the post-processing applied to the simulated waveform consists only of amplification by the effective gain of the readout chain, $G = 43$ dB. The most visible discrepancy is located on the leading edge, where the simulated signal rises more abruptly than the measured waveform due to the absence of additional frequency-domain convolution and reaches a slightly higher maximum at an earlier time. In contrast, the trailing-edge evolution is reproduced more accurately, indicating that the effective trapping description captures the dominant charge-loss behavior during the decay of the pulse. Smaller differences remain in the late-time baseline recovery. These residual deviations are consistent with the use of an effective trapping parameterization based on CCD input data, which reproduces the integral charge-loss behavior but does not explicitly resolve the microscopic distribution of trapping centers, local electric-field distortions, or grain-dependent transport paths in the material.

The measured pcCVD waveforms also show weak bipolar undershoot and overshoot after the main pulse, which are not reproduced by the simulation. They may result from the finite response of the detector–amplifier–cable–oscilloscope chain, including baseline recovery, ringing, impedance mismatch, or cable reflections. In the present simulation configuration, `TransientPropagation` calculates the detector-induced current from carrier transport and the weighting field. A complete measured impulse response or validated front-end circuit model was not included. The late-time bipolar features are therefore outside the present waveform model and are not used to evaluate the carrier-transport or trapping description.

Overall, the waveform comparison demonstrates that the implemented pcCVD model reproduces the dominant features of the measured TCT response at the field considered here. The remaining discrepancies are localized primarily in the leading-edge development and in the late trailing-edge structure, indicating that the current implementation captures the first-order transport behavior while second-order effects associated with detector-response convolution and material inhomogeneity remain outside the present description.

## 5. Conclusions and outlook

We have extended Allpix Squared with diamond-specific charge-transport models by implementing field-dependent mobility parameterizations for electrons and holes, together with an effective trapping description based on the charge collection distance (CCD). This provides a practical detector-level framework for simulating charge transport, transient signals, and charge collection in both scCVD and pcCVD diamond sensors.

For scCVD diamond, the mobility implementation was validated against published reference data using three complementary approaches. Drift velocity extracted from time-of-arrival simulations, drift velocity derived from transient times, and direct comparison of simulated and measured TCT waveforms. These mutually reinforcing validation paths show that the implemented transport models reproduce the measured carrier dynamics with good consistency over the investigated electric-field range.

For pcCVD diamond, we demonstrated that the CCD-based trapping interface reproduces the reduced charge collection and degraded transient signal shapes observed experimentally. By using measured CCD values as external inputs, the model provides a simple and experimentally accessible way to incorporate effective trapping into detector-response simulations. In this form, the implementation establishes a useful link between laboratory characterization and end-to-end detector simulation.

The present approach is intended as an effective detector model rather than a microscopic description of defect physics. Its main strength is its applicability to practical studies of signal formation, timing performance, and charge collection in diamond detectors under different operating conditions and material qualities.

Future work will focus on extending the model beyond the present effective CCD description. In particular, the inclusion of polarization effects, non-uniform internal electric fields, and spatially varying trapping in pcCVD material would further improve the realism of the simulation. Additional validation against a broader range of detector thicknesses, bias conditions, and irradiation-dependent CCD parameterizations will also be important to define the predictive range of the model for radiation-damage studies and detector optimization.

## CRediT authorship contribution statement

**Faiz Rahman Ishaqzai:** Writing – original draft, Visualization, Validation, Resources, Methodology, Investigation, Formal analysis, Data curation, Conceptualization. **Muhammed Deniz:** Writing – review & editing, Visualization, Supervision, Project administration. **Marta Baselga Bacardit:** Methodology, Data curation. **Tobias Bisanz:** Software. **Kevin Kröninger:** Writing – review & editing, Visualization, Validation, Supervision, Funding acquisition. **Jens Weingarten:** Writing – review & editing, Visualization, Validation, Supervision, Project administration, Conceptualization. **Antonia Wippermann:** Investigation, Formal analysis, Data curation.

## Declaration of generative AI and AI-assisted technologies in the manuscript preparation process:

During the preparation of this work the authors used AI in order to improve wording and sentence structure. After using these tools, the authors reviewed and edited the content as needed and take full responsibility for the content of the publication.

## Declaration of competing interest

The authors declare that they have no known competing financial interests or personal relationships that could have appeared to influence the work reported in this paper.

## Acknowledgments/funding

This work was supported by the Alexander von Humboldt Foundation.

The authors acknowledge the use of the wafer prober at TU Dortmund, funded by the Deutsche Forschungsgemeinschaft (DFG, German Research Foundation) under grant number 450639102. The authors further thank **Dustin Dobrinsky** for developing the CCD data-acquisition script, and **Simon Spannagel** and **Paul Schütze** for valuable discussions and comments on developing the simulation chain.

## Appendix A. Configuration of the diamond transport models

The following excerpt illustrates how the models implemented in this work are configured in the `TransientPropagation` module of Allpix Squared [6]. The bracketed heading identifies the module, and subsequent lines assign parameter values.

```
[TransientPropagation]
temperature = 300K
mobility_model = "diamond"
trapping_model = "pcdiamond"
pcdiamond_thickness = 500um
pcdiamond_ccd = 260um
pcdiamond_lambda_ratio = 1.3
```

The `diamond` option selects the electron PW and hole CT mobility models, while `pcdiamond` selects the CCD-based trapping model. The parameters `pcdiamond_thickness`, `pcdiamond_ccd`, and `pcdiamond_lambda_ratio` specify the sensor thickness $d$, input CCD, and trapping-length ratio $r = \lambda_h/\lambda_e$, respectively. The suffixes `K` and `um` denote kelvin and micrometers. The numerical values shown are illustrative and should be set for the detector and operating condition being simulated.

## Appendix B. Complete transient waveform simulations for mobility validation

The complete sets of simulated transient current waveforms used for the mobility validation are shown in Fig. 11. The scan contains 34 electron-field settings and 29 hole-field settings. The black markers indicate the $t_{50\%}$ crossings used to determine the FWHM, while the magenta markers indicate the time-of-arrival values obtained with `GenericPropagation`.

## Appendix C. Event selection and noise separation for the$^{90}$ sr CCD measurement

This appendix documents the separation between accepted signal events and baseline noise for the $^{90}$Sr charge spectra of Section 3.2, using the measurement at 0.4 V/μm as a representative example.

Apart from the trigger condition described in Section 3.2, the amplitude-dependent selection is the offline requirement SNR > 3 of Section 3.2.4, evaluated per waveform against the robust baseline noise $\sigma$ of that waveform.

Fig. 12(a) shows the distribution of $A/\sigma$ evaluated in the signal (coincidence) window and in a baseline-only window of equal width in the same waveforms, together with the applied cut. It can be seen that the two populations are well separated. The baseline-only distribution peaks near $A/\sigma \simeq 1.5$, with 7.1% of waveforms exceeding the applied cut of 3 and only 0.6% exceeding 5, whereas the signal distribution peaks near $A/\sigma \simeq 7$. The fraction of baseline-only windows exceeding the cut reflects the maximum being taken over ∼1500 samples and is not a contamination estimate. The corresponding bound on the contamination of the charge spectrum is given below.

Fig. 12(b) shows the same comparison in charge. The baseline-only estimator has a median of 0.32 fC, whereas all accepted events lie above 0.96 fC and the fitted most-probable value is $1.354\,\mathrm{fC} \pm 0.007\,\mathrm{fC}$, a factor of 4.2 above the noise median. Only 1.9% of the baseline-only distribution extends above the lowest accepted charge.

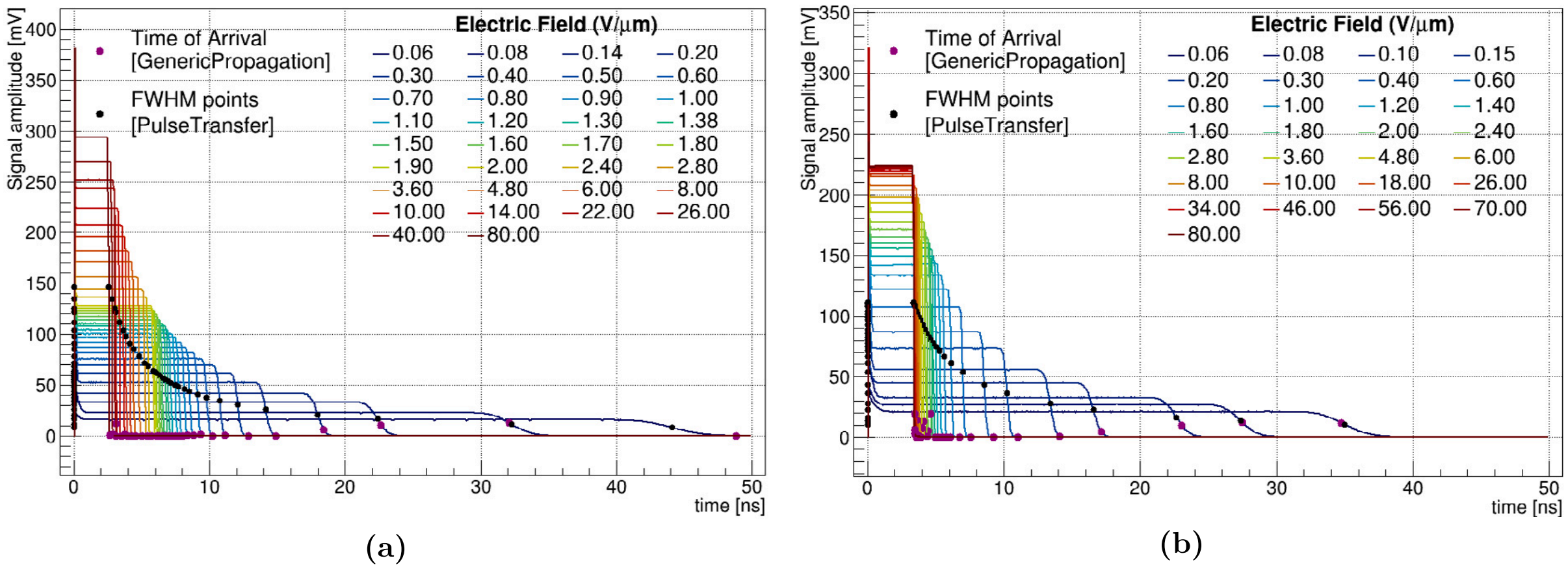

**Fig. 11.** Simulated transient current waveforms (at given electric fields) used for the extraction of transit times for (a) electrons and (b) holes. The black markers indicate the $t_{50\%}$ values on the leading and trailing edges, from which the FWHM is determined. The magenta markers represent the time-of-arrival values obtained from `GenericPropagation`.

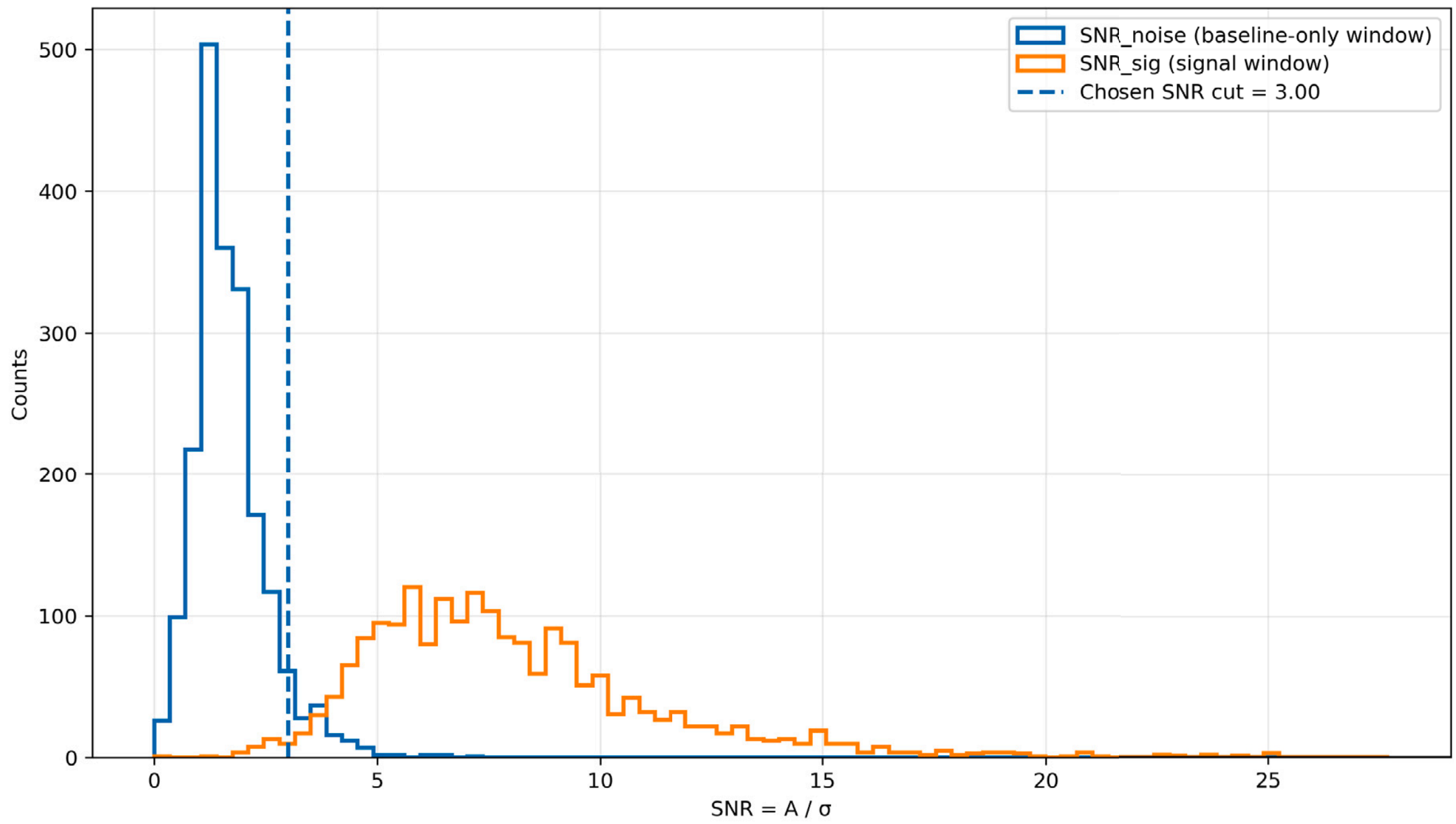

(a)

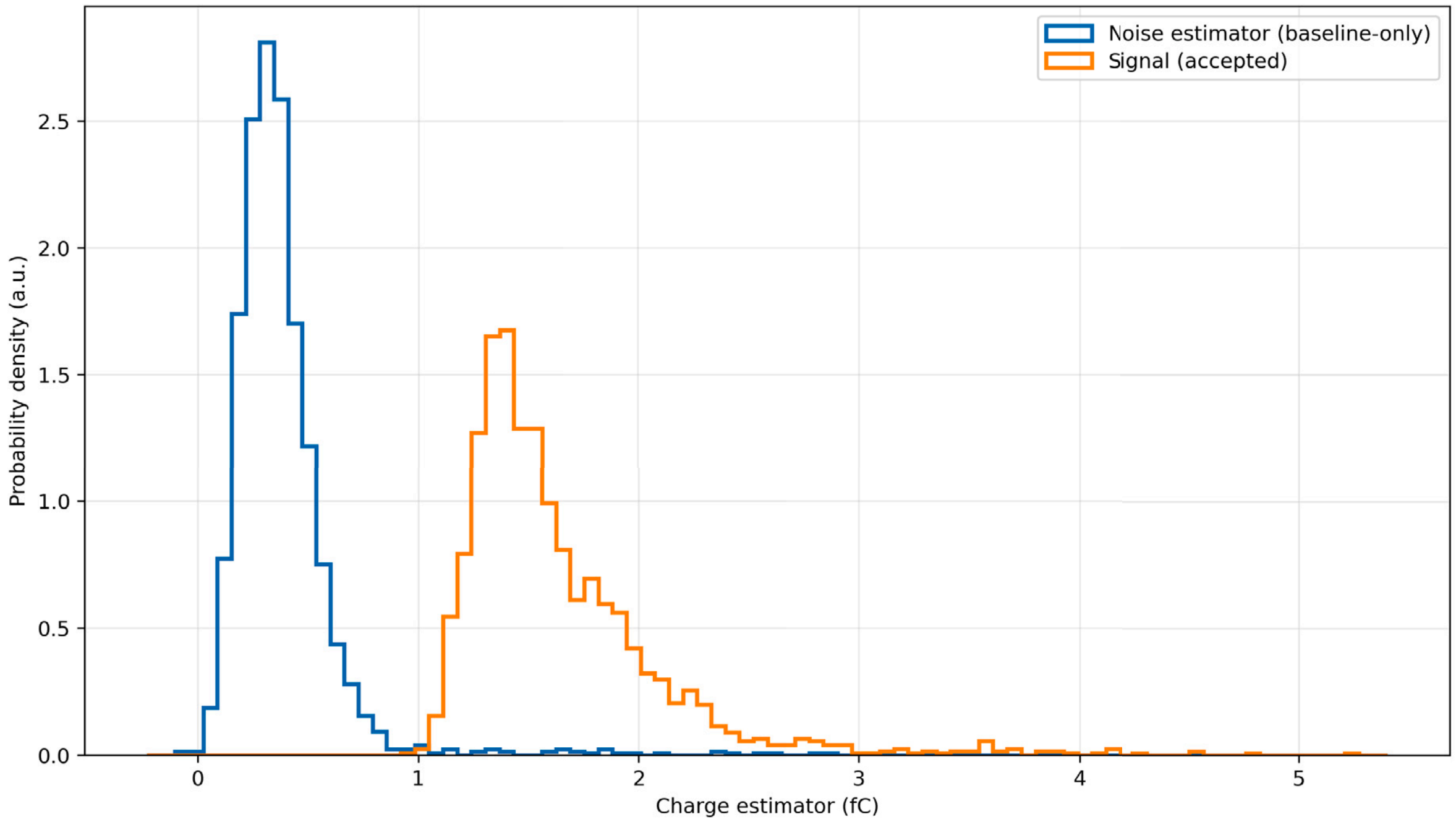

(b)

**Fig. 12.** Separation between accepted signal events and baseline noise for the $^{90}$Sr measurement at $|E| = 0.4\,\mathrm{V/\mu m}$. (a) Distribution of $A/\sigma$ evaluated in the signal (coincidence) window (orange) and in a baseline-only window of equal width in the same waveforms (blue), with the applied cut SNR $> 3$ indicated by the dashed line. (b) The same comparison expressed in charge, for the accepted events (orange) and the baseline-only window (blue), normalized to unit area.

## Data availability

The data that support the findings of this study are openly available in the TUDOdata research data repository of TU Dortmund University.

The digitized electron and hole drift-velocity and mobility data from the literature, used in Fig. are available at https://doi.org/10.17877/TUDODATA-2025-MIESBVSN.

The electrical measurement data are available at https://doi.org/10.17877/TUDODATA-2026-MXUECT.

The CCD measurement data acquired with the scintillator-trigger setup are available at https://doi.org/10.17877/TUDODATA-2026-ODJ7MD.

The TCT measurement data are available at https://doi.org/10.17877/TUDODATA-2026-LS8NHQ.

The simulation data used to validate the implementation of the mobility models in Allpix Squared using the `GenericPropagation` module are available at https://doi.org/10.17877/TUDODATA-2026-SYZQHY.

The simulation data used to validate the mobility and trapping models in Allpix Squared using TransientPropagation and PulseTransfer are available at https://doi.org/10.17877/TUDODATA-2026-F3XOXL.